\input{aipcheck}

\documentclass[
    ,final            
  ]
  {aipproc}

\layoutstyle{8x11double}

\def\gta{~\mbox{\raisebox{-.6ex}{$\stackrel{>}{\sim}$}}~}
\newcommand{\beq}{\begin{equation}}
\newcommand{\eeq}{\end{equation}}

\def\su{{\rm SUSY}}

\begin{document}

\title{QCD contributions to the thermal
history\\ of the early Universe\footnote{To appear in the Proceedings
of the Quark Confinement and the Hadron Spectrum IX
(QCHS9) Universidad Complutense de Madrid, Spain, August 30$^{th}$ - September 3$^{rd}$ 2010.}}


\author{Jose A. R. Cembranos}{
  address={William I. Fine Theoretical Physics Institute,
University of Minnesota, Minneapolis, 55455, USA \\
School of Physics and Astronomy,
University of Minnesota, Minneapolis, 55455, USA \\
Departamento de F\'{\i}sica Te\'orica I,
Universidad Complutense de Madrid, E-28040 Madrid, Spain.
}
}

\begin{abstract}
We discuss the deviations from the ideal relativistic thermal bath. These deviations are dominated from quantum chromodynamic (QCD) corrections in the most part of the parameter space of the Standard Model (SM) and the Minimal Supersymmetric Standard Model (MSSM). These effects are relevant for astrophysical precision measurements and dynamics of scalar tensor theories (SST).

\end{abstract}

\maketitle


\section{The Thermal Universe}
\label{thermo}


The thermodynamics of any substance can be characterized by its free energy density (${\cal F}$), or equivalently $g_f$, the {\it effective free energy number of relativistic degrees of freedom}:
\begin{eqnarray}
{\cal F}
&=&-g_f\,\frac{\pi^2}{90}\,T^4\,,
\label{free}
\end{eqnarray}
that is defined by normalizing to the free energy density of one non-interacting massless bosonic degree of freedom. For vanishing chemical potential, all the main properties of the fluid can be deduced directly from Eq.~(\ref{free}), such as pressure ($P$), energy density ($\rho$) or their ratio ($\omega$):
\begin{eqnarray}
P&=&- {\cal F} \,, \nonumber\\
\rho  &=& T \, \frac{\partial P}{\partial T} - P = \frac{\pi^2}{90} T^4
\left(3\,g_f+ \frac{\partial g_f}{\partial \, {\rm ln } \, T}
\right)\,, \nonumber\\
\omega &\equiv& \frac{P}{\rho} = \left( 3 + \frac{\partial \, {{\rm ln} \, g_f}}{\partial \, {\rm ln } \, T} \right)^{-1}\,.
\label{p-rho}
\end{eqnarray}
The partition function and $g_f$ can be computed diagrammatically \cite{Kapusta:1989tk}. The one loop vacuum diagram provides the free theory result. In this case, the coefficient $g_f$ is simply the number of bosonic ($N_b$) and fermionic ($N_f$) relativistic degrees of freedom: $g_{f,{\rm free}}= N_b+\frac{7}{8}\; N_f$, where the $7/8$ coefficient multiplying the fermionic contribution is due to the different value of the partition function obtained from the Fermi-Dirac thermal distribution rather than the Bose-Einstein distribution.
Higher loops account for the interactions. In \cite{copu}, it has been shown that the strong interaction is the dominant even at very high temperatures except for the case of a heavy Higgs boson. Unfortunately, the situation is not so simple since the finite temperature perturbation approach to QCD has a very slow convergence. In any case, the first order contribution has the right magnitude and the leading order correction coming from the strong interaction estimates the interaction effects in the thermal bath. In particular, for the standard model \cite{copu}:

\begin{eqnarray}
&&g_f = \frac{427}{4} - 420 \, {\tilde \alpha}_3 + {\mathcal O } \left( {\tilde \alpha}_3^{3/2} \right)
\nonumber\\
&& \frac{\partial g_f}{\partial \, {\rm ln } \, T}
= 2940 \, {\tilde \alpha_3}^2 + {\mathcal O } \left( {\tilde \alpha}_3^{5/2} \right), \nonumber\\
&& w - \frac{1}{3} = - \frac{560}{183} \, {\tilde \alpha}_3^2 + {\mathcal O } \left( {\tilde \alpha}_3^{5/2} \right)
\nonumber\\
&&
{\tilde \alpha}_3 \left(T\right) \equiv \frac{\alpha_3}{2 \pi} \simeq\frac{{\tilde \alpha}_3 \left( m_t \right)}{1+7 \, {\tilde \alpha}_3 \left( m_t \right) \, \ln\left( \frac{T}{m_t}\right)},
\label{SM}
\end{eqnarray}
with ${\tilde \alpha}_3 \left( m_t \right)\simeq 0.0172$; whereas for the MSSM \cite{copu}:
\begin{eqnarray}
&&g_f = \frac{915}{4} - 1890 \, {\tilde \alpha}_3 + {\mathcal O } \left( {\tilde \alpha}_3^{3/2} \right)
\nonumber\\
&& \frac{\partial g_f}{\partial \, {\rm ln } \, T}
= 5670 \, {\tilde \alpha_3}^2 + {\mathcal O } \left( {\tilde \alpha}_3^{5/2} \right), \nonumber\\
&& w - \frac{1}{3} = - \frac{168 \, {\tilde \alpha}_3^2}{61} + {\mathcal O } \left( {\tilde \alpha}_3^{5/2} \right)
\nonumber\\
&&
{\tilde \alpha}_3 \left(T\right)  \simeq\frac{{\tilde \alpha}_3 \left( m_\su \right)}{1+3 \, {\tilde \alpha}_3 \left( m_\su \right) \, \ln\left( \frac{T}{m_\su}\right)},
\label{MSSM}
\end{eqnarray}
Assuming for definiteness that the supersymmetric partners have masses $m_{\su} = 500 \, {\rm GeV}$, and using the running in Eq.~(\ref{SM}) from $T=m_Z$ to $T=m_\su$, we have $\alpha_3(m_{\su})\simeq 0.0153$.

We recall that the expressions (\ref{SM}) and (\ref{MSSM}) hold for temperatures greater than the masses of the particles, i.e. for $T \gta 200 \, {\rm GeV}$ in the SM, and for $T \gta 500 \, {\rm GeV}$ in the MSSM  with our assumption. At low temperatures, the effect is stronger in the SM than in the MSSM case. This is because $\alpha_3$ runs faster in the SM. However, since $\alpha_3$ decreases less in the MSSM, the departure from the noninteracting value becomes greater for the MSSM as the temperature increases.

\section{Scalar-Tensor Models}

The deviation from the ideal gas is particularly important in scalar-tensor theories (STT) of gravity, since it is not only a correction but also the main contribution to the scalar evolution. Many different types of new scalar fields are predicted different extensions of the standard model of particles and general relativity \cite{sft1,stgen}, but a standard STT can be defined by the introduction of a spin-$0$ scalar mediator in addition to the standard spin-$2$ graviton. Well known examples of such fields are the Jordan-Fierz-Brans-Dicke (JFBD) scalar which is introduced as an extension to general relativity (GR), the graviscalar (or radion) in extra dimensional models, or modili fields (as the dilaton) in the context of string theory.

The general action of these models can be written as:
\begin{eqnarray}
S &=& \int \frac{d^4 x\,\sqrt{-g}}{16 \pi G_*}  \left[ \frac{R}{A^2 \left( \phi \right)} - g^{\mu \nu} \partial_\mu \phi \,
\partial_\nu \phi - 2 U \left( \phi \right) \right]
\nonumber\\
&+& S_m \left[ g_{\mu \nu} \right]\,,
\label{act-jordan}
\end{eqnarray}
where the first term is the action for the spin-2 graviton and the scalar field, while the second term is the action for matter. This expression defines the theory in the Jordan frame, in which the standard expression of the metric $g_{\mu \nu}$ is used in the action for the matter
fields (where by ``matter'' we generally denote any field in the theory apart from $\phi$ and $g_{\mu\nu}$). However, the gravitational interaction between matter fields is modified, since the function $A \left( \phi \right)$ multiplies the Ricci scalar. Such a theory depends on two arbitrary
functions. For the sake of simplicity, we will assume that $G_*$, the bare gravitational constant, is approximately the gravitational constant that would be measured in a Cavendish-type of experiment. Under a conformal transformation
\begin{equation}
g_{*\mu\nu} = A^{-2} \left( \phi \right) \, g_{\mu \nu}\,,
\end{equation}
and for a vanishing potential ($U = 0$), the action of the system becomes
\begin{eqnarray}
S &=& \int \frac{d^4 x}{16 \pi G_*} \sqrt{-g_*} \left[ R_* - 2 g_*^{\mu \nu} \partial_\mu \phi_* \partial_\nu \phi_*  \right]
\nonumber\\
&+& S_m \left[ A^2 \left( \phi_* \right) g_{ *  \mu \nu} \right]\,,
\label{act-einstein}
\end{eqnarray}
where the scalar field $\phi_*$ is defined by
\begin{eqnarray}
\left( \frac{d \phi_*}{d \phi} \right)^2 = 3 \left( \frac{d \, {\rm ln } \, A \left( \phi \right)}{d \phi} \right)^2
 +
\frac{A^2 \left( \phi \right)}{2}
\end{eqnarray}
and $A \left( \phi_* \right)$ is short  for $A \left( \phi \left( \phi_* \right) \right) \,$.

The expression (\ref{act-einstein}) is the action of the system in the Einstein frame. It is characterized by a standard action for the spin-2 graviton; however, the combination $A^2 \left( \phi_* \right) g_{ *  \mu \nu}$, rather than the metric itself, is used in the action for matter. While the two expressions are equivalent, the Einstein frame is more often used to study the cosmology of the system (since the resulting cosmological equations are the standard ones), while the Jordan frame is more often used to study particle physics processes (since the physical lengths and masses are constant in this frame). Moreover, the Einstein frame is built in such a way that the kinetic term of the spin-2 and spin-0 mediators is diagonal so that the Cauchy problem is well-paused in this frame.

The cosmological evolution of the scalar mode is determined by the standard equation of motion
for a scalar field with an extra source term proportional to the trace of the energy momentum tensor. Assuming a perfect relatisvistic fluid form for the energy momentum tensor, its trace is simply, $\rho - 3p$, and vanishes in a radiation dominated epoch when the equation of state is characterized by $p = \rho/3$. While this is certainly a good approximation deep in the radiation era. A non-vanishing source term also arises when one includes the contribution from the trace anomaly.  Although this is generally small, typically $w - 1/3 = {\mathcal O} \left( 10^{-4} - 10^{-3} \right)$ \cite{copu}, there can be situations in which this contribution has a significant impact on the cosmology of these models. In fact, not only can it affect the evolution of the scalar field, but in some cases, it can lead to a brief early phase of contraction followed by a non-singular bounce in the cosmological scale factor. In these cases, one can also derive a maximum temperature of the universe which may provide a solution to the gravitino problem or other unwanted relics \cite{copu}.

\section{Final Remarks}

We have analyzed the consequences of deviations from the ideal relativistic thermal bath for the dynamics of scalar-tensor models. However, interaction effects are also important in the precision era of astrophysical observations that we have already entered. This fact can be illustrated in the case of a thermal relic. Although there are other possibilities \cite{other}, Dark Matter (DM) is usually assumed to be in the form of stable Weakly-interacting massive particles (WIMPs) that naturally freeze-out with the right thermal abundance. WIMPs emerge in different well-motivated particle physics scenarios as in R-parity conserving supersymmetry (SUSY) models~\cite{SUSY1,SUSY2}, universal extra dimensions (UED)~\cite{UED1, UED2}, or brane-worlds~\cite{BW1,BW2,BW3}.

The present uncertainty on the total DM density is $3\%$ \cite{Komatsu:2008hk}, and it can be improved below $1\%$ by analyzing data from the {\it Planck Surveyor}. However, in order to associate the observed precision with the thermal relic density of a particular WIMP model, it is necessary to understand the thermal bath up to the same level. Unfortunately, the uncertainties are bigger than $1\%$ percent even at high temperatures. Therefore, non-perturbative analises, as lattice studies, are necessary. Alternatively, if we are able to understand the nature of DM from other experiments as the new generation of colliders \cite{Coll}, we may have the opportunity to improve our knowledge about QCD by making precise astrophysical observations.

\begin{theacknowledgments}
This work is supported in part by
DOE grant FG02-94ER40823, FPA 2005-02327 project
(DGICYT, Spain), CAM/UCM 910309 project, and MICINN
Consolider-Ingenio MULTIDARK CSD2009-00064.
\end{theacknowledgments}

\bibliographystyle{aipproc}   

\end{document}